# Forced quantum inverted oscillator

P.A. Golovinski

Voronezh State Technical University, 84, 20-letiya Oktyabrya Street, Voronezh, 394004, Russia

*E-mail address:* golovinski@bk.ru.

New exact and asymptotic results for a quantum inverted oscillator, driven by the variable external force, are presented. To illustrate the advantages of our approach, we applied the obtained propagator to the descriptions of evolution the initial Gaussian wave packet under arbitrary time-dependent force or a $\delta$-pulse action. The alternative problem of tunneling a particle through a parabolic barrier under the influence of a low-frequency harmonic force we solve in the quasistatic approximation. Going to the Heisenberg picture in the Caldeira-Leggett model, we describe the influence of a time-dependent force on an open inverted oscillator. The resulting dynamics is presented in the form of combination the evolution of the average value of a coordinate operator and the stochastic diffusion, being induced the wave packet spreading and the Brownian motion.

## 1. Introduction

The harmonic oscillator is one of the most widely used model in theoretical physics, since it describes the behavior of an arbitrary system near a stable equilibrium state and provides foundation the quantum field theory, where free particle states are considered as boson or fermion oscillators eigenstates [1]. In this context, both classical and quantum harmonic oscillators are studied in as much detail as possible. Quite similarly, the system behavior near an absolutely unstable state, at the neighborhood of the potential energy maximum, can be described in the frame of the inverted oscillator model being universal for this type of phenomenon [2]. The solution to the quantum inverted oscillator problem gives a key to describe such different physical phenomena as the black holes evaporation [3] or the control of transmission the potential barriers in nanostructures [4].

In our work, we obtain some exact and asymptotic expressions for the dynamics of the quantum inverted oscillator under the action a time-dependent force. In Section 2, we discuss the exact solutions for the wave function and the propagator. The corresponding equations can be obtained from the formulas for the harmonic oscillator by replacing its fundamental real frequency with an imaginary value. As an example, the evolution of a wave packet experiencing arbitrary force action or $\delta$-impact is presented. In Section 3, we consider the effect of an asymptotically slowly varying harmonic force on a potential barrier penetration. We use the quasistatic approximation [5], and reduce the problem solution to the averaging the probability at a static force over the time period. In Section 4, we turn to the Heisenberg picture, which allows us to describe the above-barrier motion, simultaneously considering the external force influence and the interaction with the surrounding macroenvironment by the Langevin quantum equation, and analyzing solution we establish the deviation of the system evolution from the classical behavior.

## 2. Exact solutions

The Schrödinger equation for an inverted oscillator, perturbed by a time-dependent force $F(t)$, has the form

$$i\hbar \frac{\partial \psi}{\partial t} = -\frac{\hbar^2}{2M}\frac{\partial^2 \psi}{\partial x^2} - \frac{\Omega^2 x^2}{2}\psi - F(t)x\psi. \qquad (2.1)$$

Next, we assume the particle mass for the inverted oscillator $M=1$. Following the method [6], we introduce a new coordinate $y = x - \xi(t)$. Here, the variable $\xi(t)$ satisfies the Newton second law for an inverted oscillator

$$\ddot{\xi} - \Omega^2 \xi = F(t) \qquad (2.2)$$

corresponding the Lagrangian

$$L(\xi,\dot{\xi},t) = \dot{\xi}^2/2 + \Omega^2 \xi^2/2 + \xi F(t). \qquad (2.3)$$

By setting

$$\psi = \exp\left(\frac{i\dot{\xi}y}{\hbar}\right)\varphi(y,t), \qquad (2.4)$$

we obtain for $\varphi$ the equation

$$i\hbar \frac{\partial \varphi}{\partial t} = -\frac{\hbar^2}{2M}\frac{\partial^2 \varphi}{\partial y^2} - \frac{\Omega^2 y^2 \varphi}{2} + (\ddot{\xi} - \Omega^2 \xi - F)y\varphi - L\varphi. \qquad (2.5)$$

If we introduce a function $\chi$, such that $\varphi = \chi \exp\left(\frac{i}{\hbar}\int_0^t L\,dt\right)$, then for the new function the equation of motion takes the form

$$i\hbar \frac{\partial \chi}{\partial t} = -\frac{\hbar^2}{2}\frac{\partial^2 \chi}{\partial y^2} - \frac{\Omega^2 y^2}{2}\chi, \qquad (2.6)$$

inherent to a free inverted oscillator. Finally, the wave function of an inverted oscillator, subjected to a variable force, can be written as

$$\psi(x,t) = \chi(x - \xi(t),t) \exp\left(\frac{i}{\hbar}\dot{\xi}(x-\xi) + \frac{i}{\hbar}\int_0^t L(\xi,\dot{\xi},t)\,dt\right). \qquad (2.7)$$

A propagator for an inverted oscillator under the action force $F(t)$ can be obtained from the well-known expression for a harmonic oscillator [7] by replacing its fundamental frequency with an imaginary value $\Omega \to i\Omega$ [8] in the form

$$K(x,t \mid x_1,t_1) = \sqrt{\frac{\Omega}{2\pi i\hbar \sinh\Omega\theta}} \exp\left(\frac{i}{\hbar}S\right), \qquad (2.8)$$

where

$$\begin{aligned}S =\ &\frac{\Omega}{2\sinh\Omega\theta}[\cosh(\Omega\theta)(x^2 + x_1^2) - 2xx_1 \\ &+ \frac{2x}{\Omega}\int_{t_1}^t F(t)\sinh[\Omega(s-t_1)]\,ds + \frac{2x_1}{\Omega}\int_{t_1}^t F(t)\sinh[\Omega(t-s)]\,ds \\ &- \frac{2}{\Omega^2}\int_{t_1}^t\int_{t_1}^s F(s)F(s_1)\sinh[\Omega(t-s)]\sinh[\Omega(s_1 - t_1)]\,ds_1 ds\end{aligned} \qquad (2.9)$$

is the classical mechanics action, and $\theta = t - t_1$.

Let the initial state is prepared in the form of a wave packet [9]

$$\psi_0(x) = (2\pi\sigma^2)^{-1/4} \exp\left(-\frac{1}{4\sigma^2}(x-x_0)^2 + i\frac{p_0 x}{\hbar}\right). \qquad (2.10)$$

Such a wave packet locates the particle at the left of the barrier around the phase point $(x_0, p_0)$ with coordinate uncertainty, given by the variance $\sigma^2$. The evolution of the initial state is described by the equation

$$\psi(x,t) = \int K(x,t \mid x_1,0)\psi_0(x_1)\,dx_1. \qquad (2.11)$$

Integrating, we find the solution

$$\psi(x,t) = (2\pi\sigma^2)^{-1/4}\Gamma(t)^{-1/2}\exp\left(\frac{iS(x,t,x_0,0)}{\hbar}\right)\exp\left(-\frac{i\Omega}{2\hbar\sinh\Omega t}(x-\xi(t))^2/\Gamma(t) + i\frac{p_0 x_0}{\hbar}\right), (2.12)$$

where

$$\Gamma(t) = \cosh\Omega t + i\frac{\hbar}{2\Omega\sigma^2}\sinh\Omega t, \qquad (2.13)$$

$$\xi(t) = x_0 \cosh\Omega t + \frac{p_0}{\Omega}\sinh\Omega t + \frac{1}{\Omega}\int_0^t F(t)\sinh[\Omega(t-s)]\,ds,$$

$\xi(t)$ is a position of a classical particle at time $t$ if it started at $t = 0$ with momentum $p_0$ at a point $x_0$.

In particular, the motion, induced by the action of an extremely short pulse $F(t) = p\delta(t - t_1)$, is described by the propagator with

$$S_\delta(x,t \mid x_1,t_1) = \frac{\Omega}{2\sinh\Omega\theta}[(x^2 + x_1^2)\cosh\Omega\theta - 2xx_1] + x_1 p. \tag{2.14}$$

The wave packet, under the influence of a delta-kick, changes the initial momentum $p_0$ by an amount $p$, and it becomes equal $P = p_0 + p$. Further motion of the packet is determined exclusively by the stationary barrier of the inverted oscillator [10], so that the wave function has the form

$$\psi(x,t) = (2\pi\sigma^2)^{-1/4}\Gamma(t)^{-1/2}\exp\left(\frac{i}{\hbar}S_\delta(x,t \mid x_0,0)\right)\exp\left(-\frac{i\Omega}{2\hbar\sinh\Omega t}(x - \xi(t))^2 + i\frac{Px_0}{\hbar}\right) \tag{2.15}$$

with

$$\xi(t) = x_0 \cosh\Omega t + \frac{P}{\Omega}\sinh\Omega t. \tag{2.16}$$

The motion of the wave packet center is described by the classical dynamics, and the packet experiences a rapid spreading with time.

## 3. Tunneling mode

An opposite alternative to the short kick case is a slowly varying harmonic external force

$$F(t) = F\sin\omega_0 t \tag{3.1}$$

with frequency $\omega_0 \ll \Omega$. For a time-independent external force ($F = \text{const}$) the penetration probability for such a static parabolic barrier, i.e., the transmission coefficient

$$w \propto \exp\left(-\varepsilon(1-\beta)^2\right) \tag{3.2}$$

is given by the JWKB approximation. Here, $\varepsilon = 2\pi|E_i|/(\hbar\Omega)$ and $\beta = F/\kappa\Omega$ is the force parameter. Sequential solution of the stationary problem [11], allowing for account the existence of a reflected wave, leads to the expression

$$w = \frac{1}{1 + \exp(\varepsilon(1-\beta)^2)} \tag{3.3}$$

coinciding with equation (3.2) for $\varepsilon(1-\beta)^2 \gg 1$. The height of the potential barrier $V(\xi) = (\xi_0^2 - \xi^2)\Omega^2/2 + F(\xi_0 - \xi)$ is fall while maintaining its shape, and the transmission coefficient enlarges with increasing the magnitude $F$.

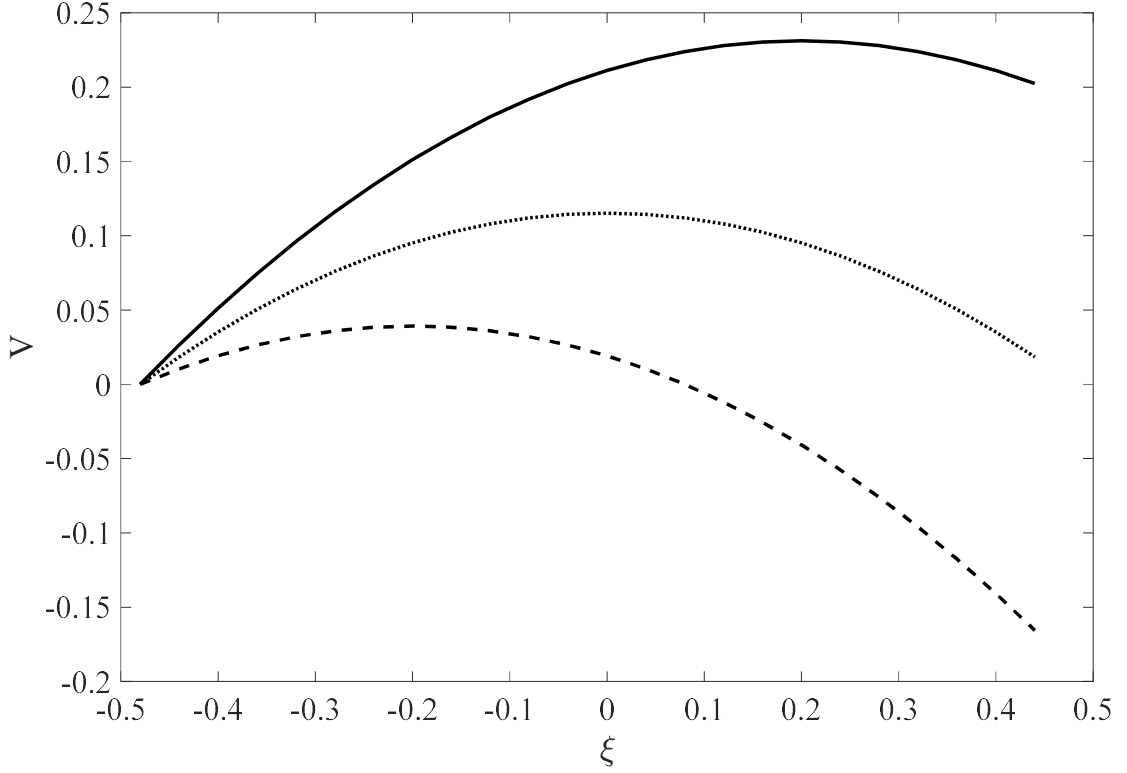

**Fig. 1.** The potential barrier form in arbitrary units: $\Omega = 1$ and $\xi_0 = -0.5$; solid line - $F = -0.2$, dotted line - $F = 0$, dashed line - $F = 0.2$.

Fig. 1 demonstrates modification of the barrier under the action of a constant force. The coordinate $\xi_0 = -\kappa/\Omega$ is the entry point for the sub-barrier motion. When the force parameter $\beta = 1$, the potential barrier is completely suppressed. As the value $\beta$ grows, the effective width of the barrier monotonously decreases, and the penetration increases.

The instant probability of a particle transmission through a barrier under the action of a time-dependent force is a function of the slowly altering parameter $\beta(t) = \beta \sin \omega_0 t$, and we find its average probability over period as

$$\overline{w} = \frac{1}{2\pi} \int_{-\pi}^{\pi} \frac{dz}{1 + \exp(\varepsilon(1 - \beta \cos z)^2)}. \tag{3.4}$$

Far from the top of the barrier, when the tunneling parameter $\varepsilon(1-\beta)^2$ is large and taking the first terms of the series expansion $\cos z = 1 - z^2/2 + \ldots$, we obtain

$$\overline{w} = A \exp(-\varepsilon(1-\beta)^2), \tag{3.5}$$

where

$$A = \frac{1}{\pi \beta} \sqrt{\frac{z}{\varepsilon}} \exp(z) K_{1/4}(z), \quad z = \frac{\varepsilon \beta^3}{16(1-\beta)}. \tag{3.6}$$

$K_{1/4}(z)$ is the Macdonald function of a fractional order [12]. The pre-exponential factor $A$ decreases the transmission coefficient compared with a constant field, similar to the manifestation a low-frequency laser field in the tunnel ionization process [5]. Fig. 2 shows the dependence of the correction factor on the force parameter for $\varepsilon = 3$.

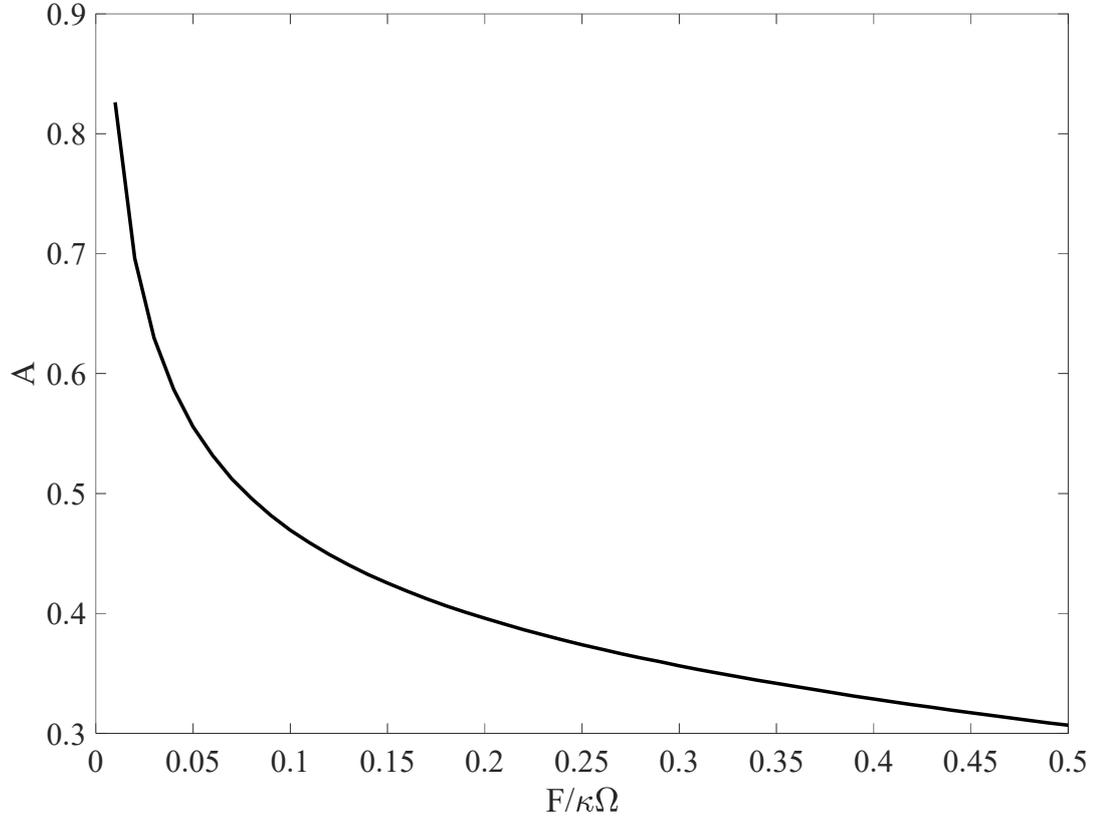

**Fig. 2.** Dependence of $A$ on the force parameter $\beta = F/\kappa\Omega$.

Our results show as far as effectively a variable force suppresses a potential barrier compared to a constant force, which should be taken into account when calculating switches with controlled nanobarriers [13].

## 4. Open inverted oscillator

In the real-world natural or artificially prepared experimental conditions, the openness of the system manifests itself, in other words, its interaction with the external environment is essential [14]. We use a microscopic model of the influence the surrounding macroenvironment, based on the idea of the linear interaction an inverted oscillator with an external heat bath, which is a set of small harmonic oscillators [15,16], known as the Caldeira-Leggett model [17-19]. The Hamiltonian of the system can be written in the form

$$H = \frac{p^2}{2} - \frac{\Omega^2 x^2}{2} + H_{bath} + H_{int} - F(t)x. \tag{4.1}$$

Hamiltonian of the heat bath and the interaction an inverted oscillator with the bath consisting of a large set of small oscillators is

$$H_{bath} + H_{int} = \frac{1}{2} \sum_j m_j \left[ p_j^2 + \omega_j^2 \left( q_j + \frac{c_j}{m_j \omega_j} x \right)^2 \right]. \tag{4.2}$$

A quantum description of an open system can be based on the Heisenberg equations of motion for the momentum and coordinate operators [20]. The momentum $p$ and the coordinate $x$ of the Brownian inverted oscillator are time-dependent operators, $p_j, q_j$ are the corresponding operators of the bath harmonic oscillators. The Heisenberg equations of motion for the Brownian oscillator have the form

$$\dot{p} = \Omega^2 x - \sum_j c_j \left( \frac{c_j}{m_j \omega_j} x + q_j \right) + F(t), \tag{4.3}$$

$$\dot{x} = p,$$

and for the bath oscillators we write down

$$\dot{p}_j = -\omega_j^2 q_j - \frac{c_j}{m_j} x, \tag{4.4}$$

$$\dot{q}_j = p_j / m_j.$$

The resulting differential equations for operators are linear, which allows them to be solved in the same way as ordinary differential equations for functions. We eliminate the bath variables by solving equations (4.4) and obtain the Heisenberg-Langevin equation of motion

$$\ddot{x}(t) - \Omega^2 x(t) + \int_0^t Z(t - t_1) \dot{x}(t_1) dt_1 = R(t) + F(t). \tag{4.5}$$

The kernel of the integral damping operator in equation (4.5) is given [21] by

$$Z(t) = \frac{2}{\pi} \int_0^\infty \frac{J(\omega)}{\omega} \cos(\omega t) d\omega, \tag{4.6}$$

where $J(\omega)$ is the bath spectral density, and the Brownian force operator

$$R(t) = -\sum_j c_j \left( q_j(0) \cos(\omega_j t) + \frac{p_j(0)}{m_j \omega_j} \sin(\omega_j t) \right). \tag{4.7}$$

Direct evaluation the function $J(\omega)$ it seems to be impossible, but the fluctuation-dissipation theorem determines the force fluctuation spectrum through the macroscopic response function and the average bath oscillation energy. Various model expressions for the response function are widely used, but in many practically important cases, the damping kernel is chosen as the exponential Drude function [16]

$$Z(t) = \gamma \omega_D e^{-\omega_D t}, \tag{4.8}$$

with spectrum

$$J(\omega) = \frac{\gamma \omega}{1 + (\omega/\omega_D)^2}. \tag{4.9}$$

The bath assumed to be in thermal equilibrium at the temperature $T$. The statistical characteristics of the variables of the bath $q_j(0)$ and $p_j(0)$ we take in accordance with the distribution for the quantum oscillator [19,22], so that

$$\langle q_j(0) q_k(0) \rangle = \frac{\langle p_j(0) p_k(0) \rangle}{m_j^2 \omega_j^2} = \frac{\hbar \coth(\hbar \omega_j / 2kT)}{2 m_j \omega_j} \delta_{jk}, \tag{4.10}$$

$$\langle q_j(0) p_k(0) \rangle = -\langle p_k(0) q_j(0) \rangle = \frac{1}{2} i \hbar \delta_{jk},$$

and $k$ is the Boltzmann constant. In the Laplace domain, the solution of equation (4.5) has the form

$$\widetilde{x}(s) = \frac{\widetilde{R}(s) + \widetilde{F}(s) + s x(0) + p(0)}{s^2 - \Omega^2 + s \widetilde{Z}(s)}. \tag{4.11}$$

For the damping kernel in the form of equation (4.8), the Laplace transform is $\widetilde{Z}(s) = \gamma \omega_D (s + \omega_D)^{-1}$. The inverse Laplace transform of equation (4.11) gives a solution in the time domain

$$x(t) = \frac{1}{2\pi i} \int_{c-i\infty}^{c+i\infty} \frac{\widetilde{R}(s) + \widetilde{F}(s) + s x(0) + p(0)}{s^2 - \Omega^2 + \gamma \omega_D s (s + \omega_D)^{-1}} e^{st} ds. \tag{4.12}$$

It is convenient to write the solution for the Brownian inverted oscillator as

$$x(t) = x(0) \dot{G}(t) + p(0) G(t) + \int_0^t G(t - t_1) [F(t_1) + R(t_1)] dt_1, \tag{4.13}$$

where fundamental solution

$$G(t) = \frac{1}{2\pi i} \int_{c-i\infty}^{c+i\infty} \widetilde{G}(s) e^{st} ds, \widetilde{G}(s) = \frac{1}{s^2 - \Omega^2 + \gamma \omega_D s (s + \omega_D)^{-1}}. \tag{4.14}$$

The function $G(t)$ satisfies the conditions $G(0) = 0, \dot{G}(0) = 1$. Since $\langle R(t) \rangle = 0$, then the average coordinate of the Brownian inverted oscillator

$$\langle x(t) \rangle = \langle x(0) \rangle \dot{G}(t) + \langle p(0) \rangle G(t) + \int_0^t G(t - t_1) F(t_1) dt_1. \tag{4.15}$$

Equation (4.15) indicates that the force fluctuations do not disturb the averaged motion of the linear inverted oscillator. Equation (4.14) gives us the system response to the step force of a single amplitude. This response is the sum of the exponential components [23] determined by residues $R_j$ at the poles $s_j$ of the function

$$\widetilde{G}(s) = \sum_{j=1}^{3} \frac{R_j}{s - s_j}. \tag{4.16}$$

The function $\widetilde{G}(s)^{-1}$ has zeros at $s = s_j$, and residues

$$R_j = \lim_{s \to s_j} \frac{s - s_j}{\widetilde{G}(s)^{-1}} = \frac{1}{d\widetilde{G}(s)^{-1}/ds\big|_{s=s_j}} = \frac{1}{2s_j + \gamma\omega_D^2(s_j + \omega_D)^{-2}}. \tag{4.17}$$

Characteristic equation for finding $r_j = s_j/\Omega$ has the form

$$r^3 + ar^2 + br - a = 0 \tag{4.18}$$

with

$$a = \omega_D/\Omega, b = \gamma\omega_D/\Omega^2 - 1. \tag{4.19}$$

Equation (4.18) has three solutions, depending on the determinant

$$D = q^2 + p^3, \tag{4.20}$$

where

$$q = \frac{a^3}{27} - \frac{ab}{6} - \frac{a}{2}, p = \frac{3b - a^2}{9}. \tag{4.21}$$

When $D > 0$, one root is real and two are complex; with $D < 0$ three roots are real, and with $D = 0$ we have one real root with threefold degeneration. Fig. 3 shows the critical line separating the parameter plane $a, b$ into two domains: with two complex and one real roots (upper) and three purely real roots (lower). The final solution for the time response function is

$$G(t) = \sum_{j=1}^{3} \frac{e^{s_j t}}{2s_j + \gamma\omega_D^2(s_j + \omega_D)^{-2}}. \tag{4.22}$$

In the quantum case, due to the possible noncommutativity of operators at different time points, a symmetrized correlation function [22]

$$\varphi(t - t') = \frac{1}{2}\langle x(t)x(t') + x(t')x(t)\rangle \tag{4.23}$$

is introduced instead of the classical correlation function. The quantum correlation function of a random force has the form [18,22]

$$\frac{1}{2}\langle R(t)R(0) + R(0)R(t)\rangle = \int_{-\infty}^{\infty} S_R(\omega)\exp(i\omega t)d\omega. \tag{4.24}$$

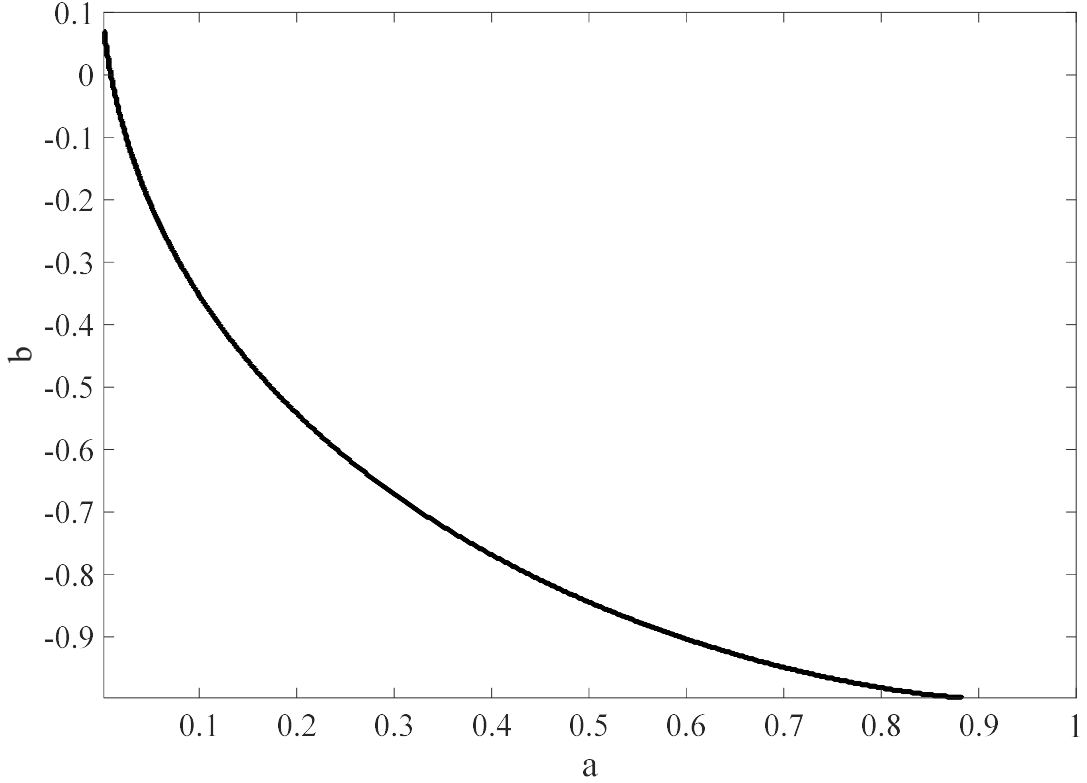

**Fig. 3.** The boundary line between domains with different determinant sign: upper - $D > 0$, lower - $D < 0$.

Here, $S_R(\omega) = \hbar J(\omega) n(\omega)/\pi$ is the spectral density of the correlation function and $n(\omega) = [\exp(\hbar\omega/kT) - 1]^{-1}$. After substitution in the equation (4.23) the coordinate operator in the form of equation (4.13), we have

$$\langle x(t)x(t') + x(t')x(t)\rangle/2 = \langle x^2(0)\rangle \dot{G}(t)\dot{G}(t') + \langle p^2(0)\rangle G(t)G(t')$$
$$+ \hat{G}(t - t_1)F(t_1)\hat{G}(t' - t_2)F(t_2) +$$
$$+ \frac{1}{2}\left[(\langle x(0)\rangle\dot{G}(t) + \langle p(0)\rangle G(t))\hat{G}(t' - t_1)F(t_1) + (\langle x(0)\rangle\dot{G}(t') + \langle p(0)\rangle G(t'))\hat{G}(t - t_1)F(t_1)\right] \quad (4.25)$$
$$+ \frac{1}{2}\left[\langle p(0)x(0)\rangle\dot{G}(t)G(t') + \langle x(0)p(0)\rangle\dot{G}(t')G(t)\right]$$
$$+ \int_{-\infty}^{\infty} d\omega S_R(\omega) e^{i\omega(t-t')} \int_0^t G(t_1) e^{-i\omega t_1} dt_1 \int_0^{t'} G(t_1) e^{i\omega t_1} dt_1.$$

Here, operator $\hat{G}$ is defined as $\hat{G}(t - t_1)F(t_1) = \int_0^t G(t - t_1)F(t_1) dt_1$. For large $t, t' \to \infty$, simultaneously greater than $\Omega^{-1}$ and a relaxation time, the last term in equation (4.25), similar to the case of harmonic oscillator [24], takes the form $\int_{-\infty}^{\infty} S_R(\omega)|\tilde{G}(i\omega)|^2 e^{i\omega(t-t')}d\omega$. The asymptotic of the quantum correlation function (4.25) contains the sum of the dynamic terms and a random force term, being generated by the spectral function, having the classical limit $kTJ(\omega)/\omega$ at $\hbar \to 0$.

The particle displacement variance, as well as for the harmonic oscillator [25], can be find in the form

$$\langle (x(t) - \langle x(t) \rangle)^2 \rangle = [\langle x^2(0) \rangle - \langle x(0) \rangle^2]\dot{G}^2(t) + [\langle p^2(0) \rangle - \langle p(0) \rangle^2]G^2(t) +$$
$$+ [\langle p(0)x(0) + x(0)p(0) \rangle - 2\langle p(0) \rangle \langle x(0) \rangle]\dot{G}(t)G(t) + \qquad (4.26)$$
$$+ \int_{-\infty}^{\infty} d\omega S_R(\omega) \left| \int_0^t G(t_1) e^{-i\omega t_1} dt_1 \right|^2.$$

For the initial state (2.11) in the form of a wave packet resting at the origin with $x_0 = 0$, $p_0 = 0$, we have

$$\langle p(0) \rangle = 0, \ \langle x(0) \rangle = 0, \qquad (4.27)$$

and $\langle p(0)x(0) + x(0)p(0) \rangle = 0$.

The oscillator response to the harmonic force (3.1) under the boundary conditions (4.27) we find as

$$x(t) = \sum_{j=1}^{3} g(s_j, t), \qquad (4.28)$$

where

$$g(s_j, t) = \frac{1}{2s_j + \gamma \omega_D^2 (s_j + \omega_D)^{-2}} \frac{F/\omega_0}{(s_j/\omega_0)^2 + 1} \left[ \exp(s_j t) - \cos \omega_0 t + \frac{s_j \sin \omega_0 t}{\omega_0} \right]. \qquad (4.29)$$

Thus, the dynamics of a particle, started from the top of an inverted oscillator potential under the action of the force (3.1), demonstrates exponential instability with the imposition of the forced harmonic oscillations, rapidly becoming invisible against the background a general increase displacement. The direction of "stalling" the particle from the equilibrium position is completely determined by the sign of $F$. Under $F = 0$ and nonzero boundary conditions, the solution (4.15) is turned to

$$x(t) = \langle x(0) \rangle \dot{G}(t) + \langle p(0) \rangle G(t) = \sum_{j=1}^{3} \frac{(\langle x(0) \rangle s_j + \langle p(0) \rangle)e^{s_j t}}{2s_j + \gamma \omega_D^2 (s_j + \omega_D)^{-2}}. \qquad (4.30)$$

For the parameters from the upper domain $a,b$, shown in Fig. 3, we have exponentials with oscillations in solution (4.30), while for the lower domain the solution is the sum of two damped and one increasing exponential functions. The appearance of oscillations is entirely depends on the time-delay the dissipative force. Comparison of equation (4.30) with equation (4.29) shows that the exponentially fast escape of a particle from an unstable equilibrium position without oscillations ($D<0$) can be compensated by the action of an external harmonic force [26]. However, such compensation is not stable in the sense of Liapunov [27].

For coordinate variance, we obtain expression

$$\left\langle (x(t)-\langle x(t)\rangle)^2 \right\rangle = \sigma^2 \dot{G}^2(t) + \frac{\hbar^2}{4\sigma^2} G^2(t) + \int_{-\infty}^{\infty} d\omega S_R(\omega) \left| \int_0^t G(t_1) e^{-i\omega t_1} dt_1 \right|^2 \quad (4.31)$$

having exactly the same form as without an external force [23]. Thermal fluctuations tend to a stationary level, while a particle exponentially fast escape from equilibrium position. As a whole, a motion can be interpreted as a change in average coordinate simultaneously controlled by internal and external forces, accompanied by the wave packet spreading. The wave packet spreading depends on the system properties itself and the effect on it the random forces, provided by the heat bath without any evidence of an applied external force.

## 5. Conclusion

We presented new analytical solutions for the wave function (2.7) and the propagator (2.8), (2.9) of the quantum inverted oscillator under the action of an external force. These solutions have the same form as for harmonic oscillator with the replacement of fundamental frequency by an imaginary value. The wave packet demonstrates the classical motion of the center of mass and the quantum spreading with time in accordance with equations (2.12), (2.13). The propagator utility is illustrated by the example of the wave packet evolution under a short pulse action. This solution can be applied to analyze the dynamics of the inverted quantum oscillator under the action of a short pulses sequence.

The effect of a low-frequency periodic force on the transmission coefficient for a parabolic potential barrier was studied in detail. The penetration of the potential barrier was evaluated by the quasistatic approximation. This made it possible to obtain an asymptotic expression for the tunneling probability (3.5) with the pre-exponential factor (3.6), which is an analogue of the ADK formula [5] for tunnel ionization of atoms in an alternating electromagnetic field. With a numerical example, we illustrate the correction to the static tunneling as a function of the force parameter.

The possibility to find exact solution to the problem of the open quantum inverted oscillator is based on quadratic form of the Hamiltonian and an exact analytical solution of the corresponding classical equations of motion. The Heisenberg-Langevin quantum equation (4.5) for an inverted oscillator in the Caldeira-Leggett model of interaction with the macroscopic environment is linear, and it is solved by the Laplace method the same manner as in classical mechanics. The solution (4.15) for an average coordinate completely coincides with the classical mechanics solution. In addition to the average motion, the initial wave packet is spread under the influence of internal force and thermal noise, regardless of the external force.

A number of issues, concerning the dynamics of the quantum inverted oscillator, still waiting to be solved. In particular, it is of interest to consider the influence of the macroscopic environment in tunneling not only for a static inverted oscillator [28], but also under the action of a variable external force. Note also, that for a nonlinear inverted oscillator, stable states can be realized due to the Kapitsa-Dirac stabilization mechanism, which quantum version [29,30] also deserves further investigation.

**Acknowledgement**

The author acknowledges support from Voronezh State Technical University. This research did not receive any specific grant from funding agencies in the public, commercial, or not-for-profit sectors.